\documentclass[twocolumn,prd,floatfix,preprintnumbers,nofootinbib,superscriptaddress, aps]{revtex4-2}
		
\usepackage{amssymb,amsmath,verbatim,mathtools,needspace,enumitem,etoolbox,graphicx,physics,microtype,afterpage,bm,soul}
\usepackage[dvipsnames]{xcolor}
\definecolor{linkcolor}{rgb}{0.0,0.3,0.5}
\usepackage[unicode, colorlinks=true, linkcolor=linkcolor, citecolor=linkcolor, filecolor=linkcolor,urlcolor=linkcolor, pdfusetitle]{hyperref}
\usepackage{orcidlink}
\usepackage[all]{hypcap}
\usepackage[T1]{fontenc}
\usepackage[utf8]{inputenc}
\usepackage{tabularx}
\usepackage{float}
  \usepackage{cancel}
\usepackage{lmodern}
\usepackage{ragged2e}
\allowdisplaybreaks
\usepackage{tikz}
\usepackage{color}
\usepackage{framed}
\usepackage{hyperref}
\hypersetup{colorlinks, citecolor=darkblue, linkcolor=black, urlcolor=darkblue}
\definecolor{rossos}{cmyk}{0,1,1,0.55}
\definecolor{bluscuro}{rgb}{0.15, 0.2, .85}
\definecolor{bluchiaro}{cmyk}{1,.3,0.,0.1}
\definecolor{ForestGreen}{rgb}{0.13, 0.55, 0.13}
\definecolor{darkblue}{rgb}{0,0, 1.39}

\newcommand{\be}{\begin{equation}}
\newcommand{\mbin}{m_\text{\tiny bin}}
\newcommand{\ee}{\end{equation}}
\renewcommand{\d}{{\rm d}}

\newcommand{\lp}{\left (}
\newcommand{\rp}{\right )}

\newcommand{\PBH}{\text{\tiny PBH}}

\def\lsim{\mathrel{\rlap{\lower4pt\hbox{\hskip0.5pt$\sim$}}
    \raise1pt\hbox{$<$}}}         
\def\gsim{\mathrel{\rlap{\lower4pt\hbox{\hskip0.5pt$\sim$}}
    \raise1pt\hbox{$>$}}}         

\begin{document}

\title{GW231123: A Possible Primordial Black Hole Origin}

\author{Valerio De Luca\orcidlink{0000-0002-1444-5372}}
\email{vdeluca@sas.upenn.edu}
\affiliation{Center for Particle Cosmology, Department of Physics and Astronomy,
University of Pennsylvania, 209 South 33rd Street, Philadelphia, Pennsylvania 19104, USA}

\author{Gabriele Franciolini\orcidlink{0000-0002-6892-9145}}
\email{gabriele.franciolini@cern.ch}
\affiliation{Department of Theoretical Physics, CERN, Esplanade des Particules 1, P.O. Box 1211, Geneva 23, Switzerland}

\author{Antonio Riotto\orcidlink{0000-0001-6948-0856}}
\email{antonio.riotto@unige.ch}
\affiliation{D\'epartement de Physique Th\'eorique and Gravitational Wave Science Center (GWSC), Universit\'e de Gen\`eve, CH-1211 Geneva, Switzerland}

\begin{abstract}
\noindent
GW231123, the heaviest binary black hole merger detected by the LIGO-Virgo-KAGRA Collaboration to date, lies in the pair-instability mass gap and exhibits unusually high component spins. In this Letter, we show that both merging black holes may have a primordial origin with smaller initial masses. The observed  masses and, crucially, the spins of GW231123 are naturally accommodated within the most vanilla primordial black hole framework, once cosmological accretion is taken into account. Interestingly, the parameter space needed to explain the inferred GW231123 rate is at the edge of the exclusion region from x-ray and CMB observations, suggesting that this interpretation can be either confirmed or ruled out. 
The upcoming O5 observing run by the collaboration should detect ${\cal O}(20)$ similar events, testing their mass-spin correlation, while next-generation detectors would be capable of observing high redshift events, as predicted in this scenario.  
\end{abstract}

\preprint{CERN-TH-2025-163} 
\maketitle

\noindent{{\bf{\em Introduction.}}} 
The LIGO-Virgo-KAGRA (LVK) Collaboration recently reported the observation of a binary black hole (BH) merger with unusually high component masses:  
$m_1 = 137^{+22}_{-17} \, M_\odot$ and $m_2 = 103^{+20}_{-52} \, M_\odot$, quoted at the $90\%$ credible level~\cite{LIGOScientific:2025rsn}.  
This detection is particularly notable because both BHs lie within the so-called ``pair instability mass gap''~\cite{1964ApJS....9..201F,1967PhRvL..18..379B,Woosley:2021xba},  
a mass range where pair-instability processes are believed to prevent BH formation through direct stellar collapse.  
In terms of mass, this is the most extreme high-significance gravitational wave (GW) event recorded to date~\cite{KAGRA:2021vkt,Wadekar:2023gea}.

\begin{figure}[ht!]
    \centering
    \includegraphics[width=0.95\linewidth]{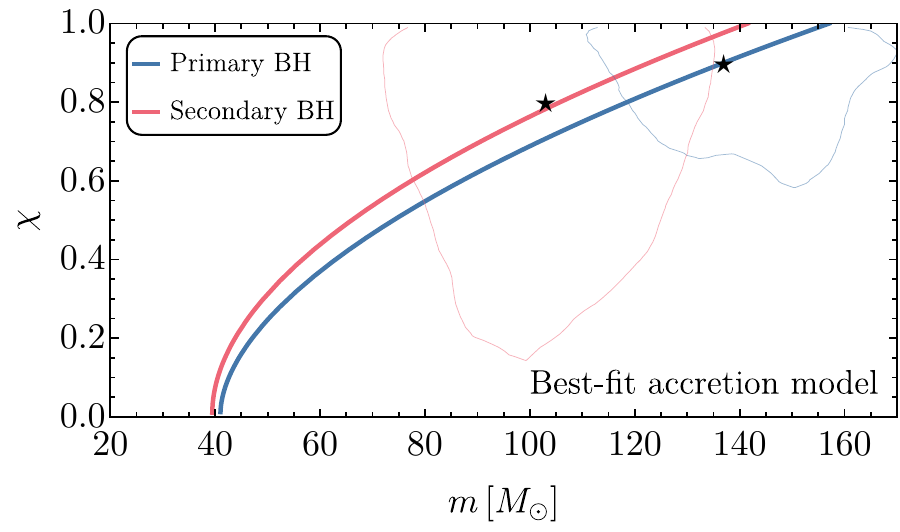}
    \caption{The contours indicate 90\% credible levels for GW231123 masses and spins. We omit the posterior at low secondary mass, as it appears only in one of the LVK analyses and does not impact our conclusions.
   The curves indicate low redshift mass versus spin in a PBH scenario (blue for primary object and red for secondary). The transition from low to high spins occurs at different masses depending on the accretion efficiency. 
   Here, we adopt a fiducial accretion model, corresponding to $z_\text{\tiny cut-off} = 24$, as explained below. This is qualitatively consistent with the inferred parameters of the binary (black stars).}
    \label{fig:spins}
\end{figure}

This system is even more massive than the previous record-holder, GW190521~\cite{LIGOScientific:2020iuh}. While several of the scenarios proposed to explain that event could also be invoked here,  
an important distinction lies in the inferred spins. In GW190521, the spin measurements were highly uncertain.  
In contrast, the spin information for the present event is remarkably well constrained. Analysis reveals a strong preference for at least one highly spinning BH,  
with dimensionless spins measured as $\chi_1 = 0.90^{+0.10}_{-0.19}$ and $\chi_2 = 0.80^{+0.20}_{-0.51}$.  
The primary BH, in particular, exhibits the highest spin ever confidently measured via GW observations~\cite{LIGOScientific:2025rsn}, see Fig.~\ref{fig:spins}.

The spin information could therefore serve as a powerful discriminator among different formation scenarios.  
For example,  hierarchical mergers offer a minimal astrophysical pathway to populate the pair-instability mass gap~\cite{Miller:2001ez,Fishbach:2017dwv,Mckernan:2017ssq, Gerosa:2017kvu, Rodriguez:2019huv}.  
However, as discussed in~\cite{LIGOScientific:2025rsn}, reproducing the large spins observed in GW231123 requires the progenitor BHs themselves to be spinning,  
a condition typically associated with large gravitational-wave recoil (or kick) velocities.  
This constrains the astrophysical environments where such hierarchical mergers could occur and be retained,  
favoring dense and massive systems with deep potential wells, such as nuclear star clusters or active galactic nuclei \cite{Antonini:2016gqe, Miller:2001ez, Antonini:2018auk, Rodriguez:2019huv, Fragione:2020nib, Mapelli:2020xeq, Sedda:2021abh, Kritos:2022non, Mahapatra:2022ngs, Bartos:2016dgn, Stone:2016wzz, Mckernan:2017ssq, Yang:2019cbr, Tagawa:2019osr, McKernan:2019beu, Vaccaro:2023cwr, Sedda:2023big}.

In this Letter, we explore whether the event may instead be compatible with a primordial black hole (PBH) origin~\cite{Zeldovich:1967lct,Hawking:1971ei,Carr:1974nx,Carr:1975qj,Bird:2016dcv,Clesse:2016vqa,Sasaki:2016jop,Eroshenko:2016hmn,Wang:2016ana,Clesse:2020ghq,Hall:2020daa,Franciolini:2022tfm,Escriva:2022bwe,Afroz:2024fzp} (see e.g.~\cite{Byrnes:2025tji,LISACosmologyWorkingGroup:2023njw} for recent reviews).  
The high spins inferred from the signal might seem at odds with standard formation of primordial binaries in a radiation-dominated era, which generally predicts low to moderate spins for binary components~\cite{Mirbabayi:2019uph,DeLuca:2019buf,Harada:2020pzb}, as assumed in \cite{Yuan:2025avq,Li:2025fnf}. Yet, such an interpretation overlooks the impact of accretion across cosmological timescales, which can significantly increase BH spins, especially in high-mass systems with $m \gg M_\odot$, such as GW231123~\cite{DeLuca:2020bjf,DeLuca:2020qqa}. Indeed, we are going to show that both observed BHs may have originated during the radiation phase of the Universe with much smaller masses. The subsequent evolution and accretion cause not only the growth of the masses, but also generate sizable spins which well explain the observed values, potentially sitting at the mass-spin correlation predicted in this scenario, see Fig.~\ref{fig:spins}.
Throughout this work, we set geometrical units, \( G = c = 1 \).

\vspace{0.1cm}
\noindent{{\bf{\em Modeling PBH evolution.}}} 
In the standard scenario, PBHs form during the radiation-dominated epoch from the collapse of large curvature perturbations, which originate from inflation~\cite{Blinnikov:2016bxu,Ivanov:1994pa,Ivanov:1997ia}. This typically happens at very high redshifts, when the relevant perturbations re-enter the Hubble sphere. A PBH of mass $m$ would form at $z_i \simeq 2 \times 10^{11} (m/M_\odot)^{-1/2}$. 
Because of the nearly spherical shape of large Gaussian overdensities~\cite{bbks} and the nature of collapse in a radiation-dominated universe, the initial dimensionless spin parameter $\chi = |J|/m^2$ is expected to be small. A characteristic estimate is~\cite{DeLuca:2019buf}
\begin{equation}
\chi_i \sim 10^{-2} \sqrt{1 - \gamma^2}\,,
\end{equation}
where $\gamma \lesssim 1$ denotes the width parameter of the curvature power spectrum.
While we will use this as a motivation, it is important to stress that we will not necessarily need to assume a specific formation scenario for the argument presented in this work, but we will only assume the initial spin of the PBHs at formation to be small. 

As the Universe evolves, PBHs may accrete baryonic matter, particularly when they reside in binary systems. In this case, accretion becomes significant if the Bondi radius of the binary exceeds its physical size~\cite{Ricotti:2007jk,Ricotti:2007au,zhang}. This condition is typically satisfied for binaries with component masses above $\mathcal{O}(10) M_\odot$, such that the binary acts as a single accreting object of mass $\mbin = m_1+m_2$. 
The mass accretion rate onto each component depends on the mass ratio $q \equiv m_2/m_1 \leq 1$, and is given by~\cite{DeLuca:2020bjf,DeLuca:2020qqa}
\begin{equation}
\dot m_1 = \dot m_\text{\tiny bin} \frac{1}{\sqrt{2(1+q)}}, \qquad
\dot m_2 = \dot m_\text{\tiny bin} \frac{\sqrt{q}}{\sqrt{2(1+q)}}\,,
\label{M1M2dotFIN}
\end{equation}
where the Bondi-Hoyle accretion rate of the binary system reads~\cite{Shapiro:1983du}
\begin{equation}
\dot m_\text{\tiny bin} = 4 \pi \lambda m_H n_\text{\tiny gas} v_\text{\tiny eff}^{-3} \mbin^2\,,
\label{R1bin}
\end{equation}
with $m_H$ the hydrogen mass, and $n_\text{\tiny gas}$ the cosmic mean gas density. The parameter $\lambda$ captures a variety of physical effects, including gas viscosity, Hubble expansion, and Compton coupling with CMB photons~\cite{Ricotti:2007jk}.
Interestingly, we can see from Eq.~\eqref{M1M2dotFIN} that, since the less massive component in a binary accretes more efficiently, evolution of masses tends to increase the mass ratio $q$ over time according to
\begin{equation}
\dot q = q \left( \frac{\dot m_2}{m_2} - \frac{\dot m_1}{m_1} \right) > 0\,.
\end{equation}
Therefore, a primordial interpretation of GW231123 with spins induced by accretion would be incompatible with a small mass ratio.

In this work, we follow~\cite{DeLuca:2020bjf} and the references therein, and include an important enhancement in the accretion rate which arises due to the presence of dark matter (DM) halos around PBHs, especially when PBHs make up only a subdominant component of the DM~\cite{Ricotti:2007au,Adamek:2019gns,Mack:2006gz}. This catalytic effect is incorporated into the parameter $\lambda$~\cite{DeLuca:2020bjf,DeLuca:2020bjf}.
The fact that PBHs contribute to a subdominant DM component is a good assumption given the existing bounds on the PBH abundance in the solar mass range \cite{Carr:2020gox}, which is also confirmed a posteriori, given the value of the abundance $f_\text{\tiny PBH}$ needed to match the merger rate of GW231123. 

Given that the accretion efficiency drops significantly around the onset of structure formation~\cite{Hasinger:2020ptw,Hutsi:2019hlw,Ali-Haimoud:2017rtz}, we introduce a redshift cut-off $z_\text{\tiny cut-off}$ beyond which we neglect further accretion. This benchmark value is motivated by detailed studies~\cite{DeLuca:2020bjf,DeLuca:2020qqa} and accounts for various model uncertainties including X-ray preheating~\cite{Oh:2003pm}, astrophysical feedback (both local and global)~\cite{Ricotti:2007au,Ali-Haimoud:2016mbv, Facchinetti:2022kbg}, and mechanical feedback~\cite{Bosch-Ramon:2020pcz}.
We neglect the possible role of radiative feedback as proposed in the Park-Ricotti model~\cite{Park:2010yh,Park:2011rf,Park:2012cr,Sugimura:2020rdw,Scarcella:2020ssk}, which accounts for the impact of trapped radiation on the surrounding medium, leading to a localized enhancement in the speed of sound within the ionized region. Such an effect is expected to reduce the accretion efficiency. It should be emphasized that an $\mathcal{O}(1)$ change in the BH mass is sufficient to induce a complete spin-up, provided that accretion proceeds through a disk geometry. 
Overall, it is important to stress that large uncertainties on the PBH accretion model remain \cite{DeLuca:2023bcr,Serpico:2024cdz,Jangra:2024sif}. We expect that the parameter $z_\text{\tiny{cut-off}}$ used here could be traded for more physical quantities that encode the uncertainties in the accretion efficiency for PBHs in binaries~\cite{Scarcellainprep}.

Accretion has several important consequences on the PBH population. It increases the average mass, therefore shifting the distribution toward higher masses, developing a high-mass tail~\cite{DeLuca:2020bjf}. Additionally, accretion alters the PBH abundance fraction in DM as a function of the redshift~\cite{DeLuca:2020fpg} 
\begin{equation}
f_\PBH (z) = \frac{\langle m (z) \rangle}{\langle m (z_i) \rangle (f_\PBH^{-1}(z_i) - 1) + \langle m(z) \rangle}\,,
\label{fPBHevo}
\end{equation}
where $\langle m (z) \rangle$ is the average PBH mass at redshift $z$ and $f_\PBH^{-1}(z_i)$ is the PBH abundance at formation. This redshift dependence is crucial when comparing theoretical predictions with observational constraints.

The angular momentum carried by the infalling gas plays a key role in shaping the PBH spin evolution~\cite{Berti:2008af}. In a regime of efficient baryonic accretion, typically realized for mildly super-Eddington rates, a thin accretion disk is expected to form in the equatorial plane of the PBH, leading to efficient transfer of angular momentum from the baryonic matter to the object, which eventually spins up~\cite{DeLuca:2020bjf,DeLuca:2020qqa}. In this thin-disk regime, the spin evolution can be described using a geodesic accretion model, whose corresponding evolution equation reads~\cite{Bardeen:1972fi,Brito:2014wla,Berti:2008af,DeLuca:2020qqa}
\begin{equation}
\dot \chi_j = g(\chi_j) \frac{\dot m_j}{m_j}\,,
\end{equation}
where $\chi_j$ is the dimensionless spin of the $j$-th PBH and $g(\chi_j)$ is a dimensionless growth function set by the properties of the disk accretion model and shown explicitly in~\cite{DeLuca:2020qqa}. We emphasize that $g(\chi_j)$ provides a physically motivated but model-dependent parametrization of spin growth under efficient thin-disk accretion, capped at $\chi_\text{\tiny max} = 0.998$ due to radiation losses~\cite{thorne,Gammie:2003qi}.
A phenomenological fit of the mass-spin relation obtained using this accretion model is reported in Ref.~\cite{Franciolini:2021xbq}, which is shown in Fig.~\ref{fig:spins}.
This mass-spin positive correlation is similar, but not entirely degenerate, with the one expected from astrophysical hierarchical merger scenarios, although the current limited GWTC-3 dataset cannot distinguish between the two \cite{Franciolini:2022iaa}.
We emphasize that here accretion occurs at high redshifts, when the binaries are still widely separated, so the orientations of the individual PBH spins remain uncorrelated.

\vspace{0.1cm}
\noindent{{\bf{\em Merger rate of primordial GW231123-like events.}}}
The initial abundance and distribution of PBHs at formation controls the probability that a pair of PBHs will decouple from cosmic expansion and form a gravitationally bound binary in the early Universe. Such decoupling occurs due to gravitational interactions that overcome the Hubble flow. Once a binary is formed, the surrounding matter, including other PBHs and small-scale density perturbations, exerts torques on the system, shaping its initial semimajor axis and orbital eccentricity (see e.g. \cite{Raidal:2024bmm} for a review).

The resulting PBH binary merger rate, in the absence of accretion effects, is given by~\cite{Raidal:2018bbj,Vaskonen:2019jpv,Franciolini:2022ewd}
\begin{align}
\label{mergerrate}
\frac{\d R}{\d m_1^i \d m_2^i} 
= &
\frac{1.6 \cdot 10^6}{{\rm Gpc^3 \, yr}} f_\PBH^{\frac{53}{37}} (z_i)
\eta_i^{-\frac{34}{37}}
\left( \frac{t}{t_0} \right)^{-\frac{34}{37}}  
\left( \frac{m^i_\text{\tiny bin}}{M_\odot} \right)^{-\frac{32}{37}} 
\nonumber \\
\times &S\left( m^i_\text{\tiny bin}, f_\PBH (z_i), \psi\right)
\psi(m^i_1, z_i) \psi (m^i_2, z_i)\,,
\end{align}
where $\mu^i = m^i_1 m^i_2 / m^i_\text{\tiny bin}$, $\eta_i = \mu^i / m^i_\text{\tiny bin}$ and $\psi (m_j^i, z_i)$ denote the initial reduced mass, symmetric mass ratio, and mass function, respectively. The time variable $t$ denotes the merger time, and $t_0$ is the current age of the Universe.
The suppression factor $S\left(m^i_\text{\tiny bin}, f_\PBH, \psi\right)$ accounts for  the evolution of the binary semimajor axis and eccentricity due to binary tidal torques and interactions with the surrounding environment in both the early- and late-time Universe.
This can be written as the product of two separate contributions,
$S \equiv S_{\rm E} \times S_{\rm L}$.
The former $S_{\rm E}$ reduces the PBH merger rate due to interactions, close to the formation epoch, between the forming binary and both the surrounding DM inhomogeneities, with characteristic variance $\sigma_\text{\tiny M}^2 \simeq 3.6 \times 10^{-5}$, as well as neighboring PBHs~\cite{Ali-Haimoud:2017rtz,Raidal:2018bbj,Liu:2018ess}.
It is calibrated against dedicated numerical simulations~\cite{Raidal:2018bbj} and fully captures the relevant physics of early-Universe binary assembly.
The latter contribution, $S_{\rm L}$, accounts for the effect of successive disruption of binaries that populate PBH clusters formed from the initial Poisson inhomogeneities. It is conservatively estimated by assuming that the entire fraction of binaries residing in dense environments is disrupted \cite{Vaskonen:2019jpv}. This suppression is negligible for small $f_\PBH \lesssim {\cal O}(10^{-2})$. (We will only consider here formation scenarios where initial clustering is negligible~\cite{Desjacques:2018wuu,DeLuca:2020jug,DeLuca:2022uvz,Crescimbeni:2025ywm}, and PBH structure formation proceeds through Poisson-induced inhomogeneities~\cite{Carr:2018rid,Inman:2019wvr,DeLuca:2020jug}.)
Moreover, the fraction of PBHs involved in mergers remains small, approximately $\mathcal{O}(10^{-2} f_\PBH^{16/37})$~\cite{Liu:2019rnx,Wu:2020drm}, implying that second-generation mergers are negligible in the LVK observational window~\cite{DeLuca:2020bjf,Liu:2019rnx,Wu:2020drm}. 

When baryonic accretion is included, the binary merger rate is enhanced due to the growth in both the total and reduced masses, as well as increased orbital hardening from energy loss via GW emission~\cite{DeLuca:2020qqa,Caputo:2020irr,Peters:1963ux,Peters:1964zz}. 
We model this phase assuming that mass accretion proceeds adiabatically, i.e. on timescales much longer than the orbital period. In this regime, the orbital eccentricity remains a constant of motion during the accretion-driven evolution, while the evolution of the binary semimajor axis leads to a gradual shrinking of the orbit as the PBH masses increase. This accretion-driven inspiral enhances the probability of merger, resulting in a corrected merger rate of the form~\cite{DeLuca:2020qqa}
\begin{align}
\label{mergerrateacc}
\d R_\text{\tiny  acc} 
&= \d R  
\left( \frac{ m_\text{\tiny bin}(z_\text{\tiny cut-off})}{m_\text{\tiny bin} (z_i) } \right)^{\frac{9}{37}} 
\left( \frac{\eta (z_\text{\tiny cut-off})}{\eta (z_i) } \right)^{\frac{3}{37}} \nonumber \\
& \quad \times  
\exp\left[ \frac{12}{37} 
\int_{t(z_i)}^{t(z_\text{\tiny cut-off})} \d t 
\left( \frac{\dot m_\text{\tiny bin}}{m_\text{\tiny bin}} + 2 \frac{\dot \mu}{\mu} \right) \right]\,.
\end{align}
The integration extends from the formation time $t(z_i)$ to the cut-off time $t(z_\text{\tiny cut-off})$, after which accretion is assumed to become inefficient.

We note that the merger-rate expression above follows the standard framework for early-Universe PBH binary formation, and several potential sources of uncertainty exist, including the PBH mass function, details of binary assembly, and possible initial clustering. However, for the low PBH abundances and narrow mass range relevant to the GW231123 event,  variations in the initial PBH properties have a negligible effect on the predicted merger rate. Similarly, clustering effects are minimal in this regime, and late-time dynamical channels are subdominant. As a result, while uncertainties exist in principle, the merger-rate predictions in the regime of interest are robust and not meaningfully affected.

\vspace{0.1cm}
\noindent{{\bf{\em Results.}}}
In order to match the inferred mass and spin parameters, we find a representative preference for $z_\text{\tiny cut-off}  = 24$. Using this reference value, we find that the initial masses of GW231123 were
\begin{equation}
    m^i_1 = 85.5 M_\odot\,,
    \quad 
    m^i_2 = 59.1 M_\odot\,.
\end{equation}
We emphasize that this value of $z_\text{\tiny cut-off}$
is not uniquely selected: similar agreement with the observed masses and spins can be obtained for slightly heavier (lighter) initial PBH masses by adopting slightly higher (lower) cutoff redshifts, reflecting a degeneracy between these parameters. A more quantitative determination of this degeneracy and of the posterior distribution for the cutoff redshift would require a dedicated Bayesian analysis, which we leave for future work.
In Fig.~\ref{fig:mass,spin,evo}, we show the mass and spin evolution as a function of the time until the cut-off epoch, at which we assume accretion becomes ineffective.

\begin{figure}[t!]
    \centering
    \includegraphics[width=1
    \linewidth]{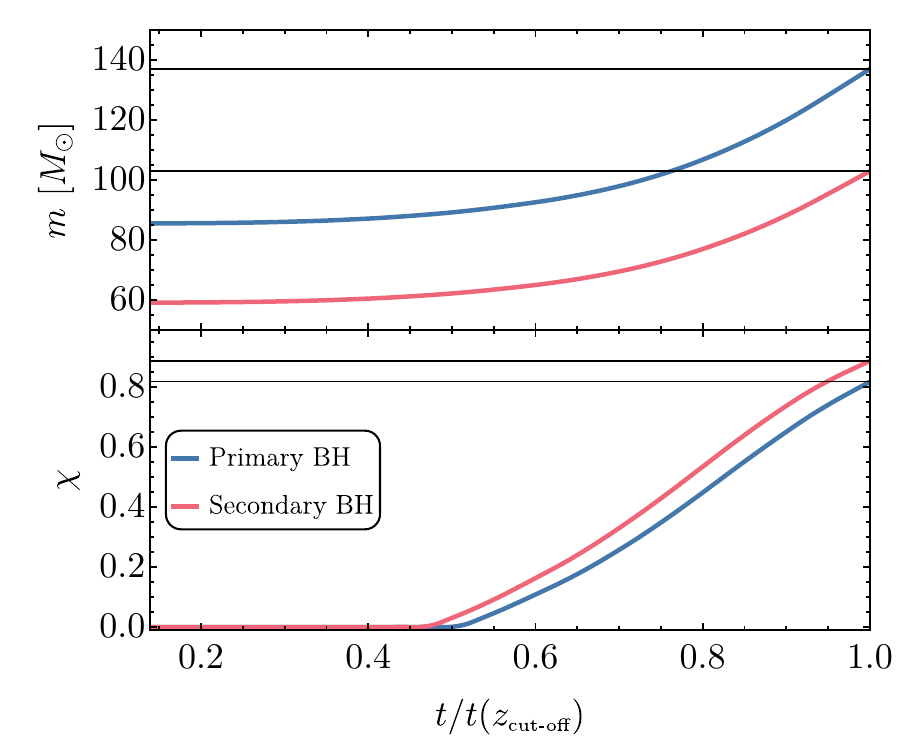}
    \caption{Time evolution of the PBH masses (top) and spins (bottom) of the GW231123 event. The black solid lines indicate the measured central values.}
    \label{fig:mass,spin,evo}
\end{figure}

A widely used, model-independent representation of the PBH mass distribution at formation is the log-normal function
\begin{equation}
\psi(m,z_i) = \frac{1}{\sqrt{2\pi}\sigma  m} \exp \left[ -\frac{\log^2 (m/M_*)}{2 \sigma^2} \right]\,,
\label{psi}
\end{equation}
which is characterized by a reference mass scale $M_*$ and a width $\sigma$. This form provides a robust description of sharply peaked PBH mass spectra and effectively captures the outcome of a broad class of PBH formation mechanisms, including those arising from symmetric peaks in the primordial power spectrum~\cite{Dolgov:1992pu,Carr:2017jsz}. It therefore serves as a phenomenological model for an initial PBH mass function.
To illustrate consistency with GW231123, we assume the mass distribution to be peaked at $M_* = 73 M_\odot$ and width $\sigma =0.2$. 
We emphasize that these parameters are not uniquely determined: similar agreement with the event's observed properties can be obtained by varying the central mass and width within a plausible range. Our choice serves to show that this event could be embedded in an explicit PBH scenario, while more general or broader mass functions could also be considered (see e.g. \cite{Franciolini:2022tfm,Escriva:2022bwe}). Using the representative values above, we compute the merger rate following Eq.~\eqref{mergerrateacc}. 
We note that the lognormal distribution with the parameters chosen above describes the portion of the PBH population contributing to the heavy mass range observed by LVK.

\begin{figure}[t!]
    \centering
    \includegraphics[width=1\linewidth]{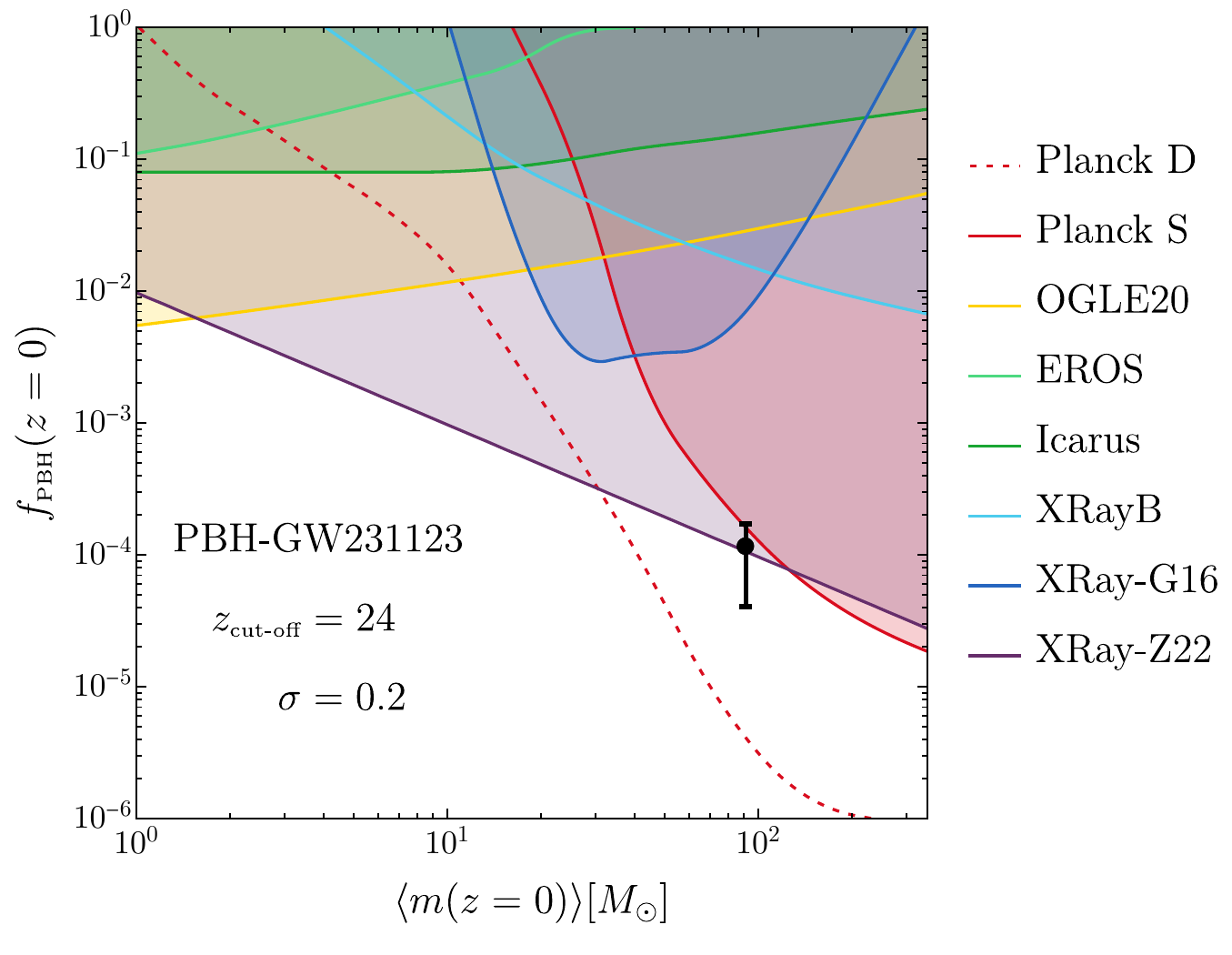}
    \caption{
    Constraints on the current mass and abundance of a PBH population assuming an initial width of $\sigma = 0.2$ and the benchmark accretion scenario with $z_\text{\tiny cut-off} = 24$. 
    The black uncertainty interval corresponds to the range of $f_\PBH(z=0)$, which matches the inferred  90\% C.I. of the merger rate density of GW231123. We superimpose other constraints described in the text. }
    \label{fig:main}
\end{figure}

In Fig.~\ref{fig:main}, we show the resulting parameter space for the late-time mass and abundance needed to explain the merger rate observed for the event GW231123, given by $R = 0.08^{+0.19}_{-0.07} \, {\rm Gpc}^{-3} {\rm yr}^{-1}$~\cite{LIGOScientific:2025rsn}; see also Ref.~\cite{rate}. In this calculation, we fix the component masses, redshift, and PBH mass function associated with the event, so that matching the predicted merger rate of Eq.~\eqref{mergerrateacc} to the observed rate then uniquely determines $f_\PBH$.
The figure thus directly illustrates the abundance and mass requirements for GW231123 within the PBH framework.

In the mass range of interest for our discussion, the most relevant constraints come from CMB anisotropies produced by accreting PBHs in the early Universe~\cite{Ali-Haimoud:2016mbv, Serpico:2020ehh}.
This bound constrains the PBH population in the redshift range $z
\sim (300\divisionsymbol 600)$, and thus is sensitive to the PBH mass function before accretion took place. It is computed assuming spherical (S) and disk (D) geometry of accretion at high redshifts~\cite{disk}.
Other constraints come from comparing the {\it late time} emission of electromagnetic signals from interstellar gas accretion onto PBHs with observations of galactic radio and x-ray isolated sources~(XRay-G16)~\cite{Gaggero:2016dpq,Manshanden:2018tze} and 
x-ray binaries~(XRayB)~\cite{Inoue:2017csr},
x-ray and radio backgrounds (XRay-Z22)~\cite{Ziparo:2022fnc}, lensing searches of massive compact halo objects toward the Large Magellanic Clouds (EROS)~\cite{Allsman:2000kg}, fast transient events near critical curves of massive galaxy clusters (ICARUS)~\cite{Oguri:2017ock}, and observations of stars in the Galactic bulge by the Optical Gravitational Lensing Experiment (OGLE 20)~\cite{Niikura:2019kqi,Mroz:2024mse}.
The bounds are shown accounting for the accretion history as discussed in~\cite{DeLuca:2020fpg}.

Figure~\ref{fig:main} indicates that the representative scenario supporting the interpretation of GW231123 as a PBH event necessitates model parameters that lie near the boundary of the exclusion regions imposed by x-ray and CMB observations. This proximity suggests that forthcoming or improved GW-independent observations could play a crucial role in either confirming or ruling out the PBH interpretation of this GW event.

Assuming the narrow mass function and representative parameters required to explain GW231123 as a PBH merger, one can make concrete forecasts for future GW observations. The upcoming O5 observing run by the LVK Collaboration  \cite{KAGRA:2013rdx}, as well as next-generation detectors such as the Einstein Telescope (ET) \cite{Punturo:2010zz, Hild:2010id, Maggiore:2019uih,Branchesi:2023mws} and Cosmic Explorer \cite{Reitze:2019iox, Evans:2021gyd, Evans:2023euw}, are expected to substantially enhance sensitivity to a wider range of source parameters. These improvements will enable more precise measurements of the BH population, offering a robust test of this PBH interpretation across a broader range of masses and redshifts \cite{Franciolini:2023opt}.
One consequence of the primordial interpretation is that additional high-mass BH mergers similar to GW231123 should exhibit the mass-spin correlation predicted in this scenario, see Fig.~\ref{fig:spins}. Our representative PBH mass function predicts at least $\mathcal{O}(20)$ such events in the LVK O5 run, in addition to any astrophysical contribution, with similar rates expected for other viable parameter choices. The  uncertainties in GW231123's source properties  allow for a range of parameter choices, but these lead to qualitatively similar predictions for both event rate and mass-spin correlation. Multiple observations would test whether a single accretion model can simultaneously explain the ensemble: inconsistent accretion parameters across events would indicate model limitations or astrophysical origins, while consistency, combined with the electromagnetic signatures discussed below, would strengthen this interpretation. Observing significantly fewer events, or events inconsistent with the predicted mass-spin correlation, would disfavor this PBH interpretation.

\begin{figure}[t!]
    \centering
    \includegraphics[width=1\linewidth]{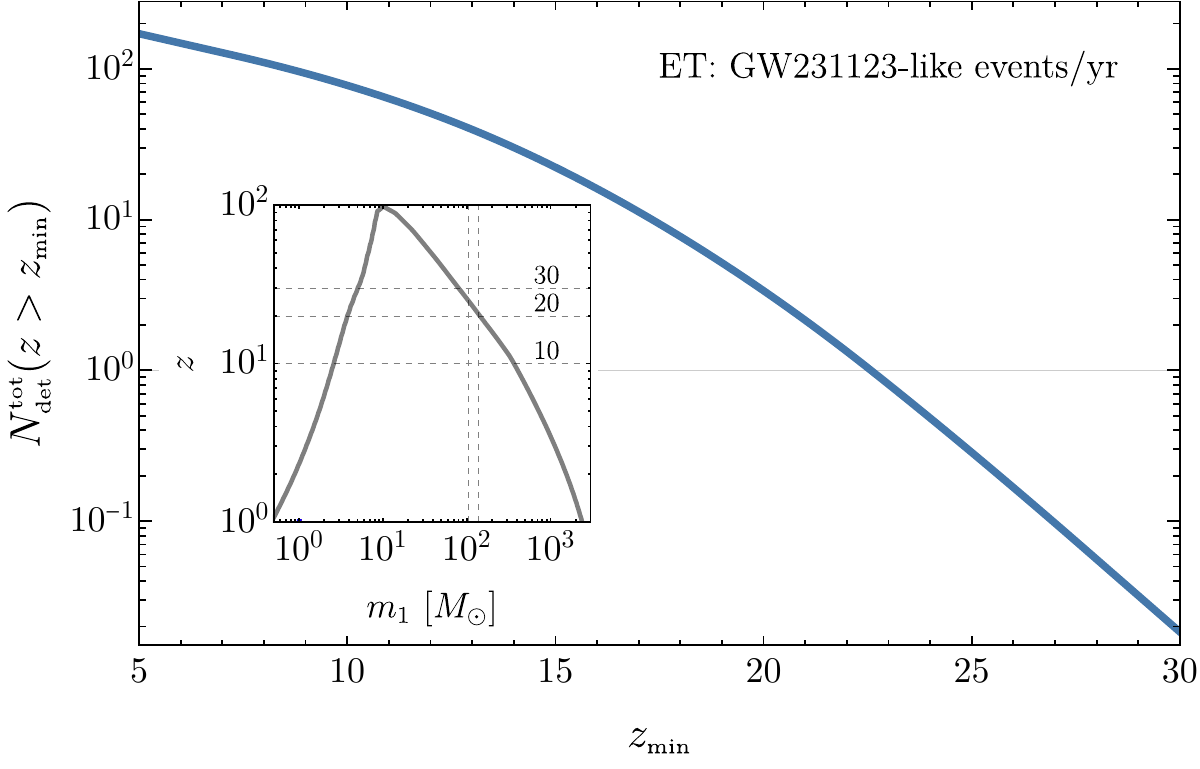}
    \caption{Prediction for the number of high-redshift GW231123-like events detectable by ET per year of observations. The inset shows the ET redshift horizon, with the dashed vertical lines indicating the GW231123 masses to guide the eye. }
    \label{fig:ET}
\end{figure}

A second prediction concerns the detection of mergers at high redshifts, which would be extremely challenging to explain through conventional astrophysical channels, but are a distinctive feature of PBH formation scenarios due to their early-Universe origin~\cite{Koushiappas:2017kqm,DeLuca:2021hde, Franciolini:2021xbq}.
Assuming the predicted sensitivity of ET, we can provide an estimate of the number of high-redshift detectable events expected within the PBH hypothesis, $N^\text{\tiny tot}_\text{\tiny det} (z> z_\text{\tiny min})$~\cite{DeLuca:2021hde}.
This prediction, shown in Fig.~\ref{fig:ET} as a function of source redshift, indicates that we expect about ten events similar to GW231123 above $z \gtrsim 15$, but none much beyond $z \sim \mathcal{O}(20)$. The rapid drop-off of observable events arises from the sharp decrease of the ET sensitivity reach at higher redshifts for GW231123 masses, as can be appreciated from the inset of the figure. 
While an astrophysical foreground of mergers at this intermediate redshift can be produced by Pop III stars \cite{10.1093/mnras/207.3.585,Bromm:2003vv,Belczynski:2004gu,Tanikawa:2021qqi,Hijikawa:2021hrf}, population studies focusing on the merger rate evolution at high redshifts would be able to disentangle the two populations \cite{Ng:2022agi}. These observations, if realized, would provide critical evidence to support or falsify the PBH interpretation of GW231123.
Let us also mention that primordial GW231123-like events are expected to have negligible eccentricity, since the binary circularizes efficiently during its earlier evolution, different from hierarchical astrophysical mergers~\cite{Zevin:2021rtf}. Its measurability improves with the observation of the low frequency part of the signal, thus making LISA an option~\cite{LISA:2017pwj,LISA:2024hlh}. However, even if the binary undergoes its early inspiral in the LISA sensitivity band, it is characterized by a low signal-to-noise-ratio~$\sim \mathcal{O}(5)$, thus suggesting that a measurement of the eccentricity seems unlikely~\cite{Franciolini:2021xbq}.

Finally, we note that our PBH scenario requires the formation of accretion disks to spin up the PBHs, and thus inevitably predicts tidal interactions between the accretion disks surrounding each PBH (assuming that they survive interactions with external perturbations), leading to short-lived tidal disruption events and to resulting multi-messenger electromagnetic emission. This process is expected to occur once the binary’s semimajor axis approaches the Roche radius, at which point the disks are disrupted and produce luminous flares.

We can roughly estimate the corresponding spectral density by considering a simplified model in which each PBH is surrounded by an accretion disk of effective size $L \equiv m\,\tilde{L}$---characterizing, for instance, the radius of maximum density---and effective mass $m_{\tiny\rm env} \equiv \epsilon\, m \ll m$.
For an approximately equal-mass binary, tidal disruption occurs when the separation reaches the Roche radius $d_\text{\tiny Roche} = \gamma\, m\,\tilde{L}$~\cite{DeLuca:2024uju, Cannizzaro:2024fpz}, where $\gamma$ ranges from $1.26$ for rigid bodies to $2.44$ for fluid bodies~\cite{Shapiro:1983du}. Assuming that the energy released during disruption is of order $f_\text{\tiny rad} \, (m\, m_\text{\tiny env} / d_\text{\tiny Roche})$, where $f_\text{\tiny rad}$ is an efficiency factor encoding the fraction of energy emitted as visible standard model radiation, the corresponding power output over a single orbital period can be estimated as~\cite{Berti:2024moe}
\begin{align}
\mathcal{P} \simeq 10^{14} L_\odot f_\text{\tiny rad} \left(\frac{\tilde L}{10^2} \right)^{-5/2} \left(\frac{\epsilon}{10^{-3}} \right)^{3/2} \,,
\end{align}
where $L_\odot = 3.8 \times 10^{26} \, {\rm W}$ is the luminosity of the Sun, and we have assumed values for the disk parameters compatible with realistic predictions for astrophysical thin disks, e.g. $\tilde L \in (10 \divisionsymbol 100)$~\cite{Abramowicz:2011xu, Inayoshi:2019fun} and $\epsilon \lesssim 10^{-3}$ to avoid precession effects~\cite{Tiede:2023cje}. 

To evaluate the detectability of this radiation with a ground-based telescope, we take an angular resolution of $\delta\Omega = 1^\circ \times 1^\circ = (\pi/180)^2\,{\rm sr}$ and place the binary at a distance $d_\text{\tiny L} \sim \mathcal{O}(1)\,{\rm Gpc}$ at the time of tidal disruption. Under these assumptions, the frequency-weighted spectral density becomes~\cite{Berti:2024moe}
\begin{align}
\frac{\mathcal{P}}{d_\text{\tiny L}^2 \delta \Omega} & \simeq 10^{-13} \frac{\rm W}{\rm m^2 sr}   \lp \frac{f_\text{\tiny rad}}{10^{-6}} \rp \lp \frac{d_\text{\tiny L}}{\rm Gpc} \rp^{-2}
\nonumber \\
& \times \left(\frac{\tilde L}{10^2} \right)^{-5/2} \left(\frac{\epsilon}{10^{-3}} \right)^{3/2}\,,
\end{align}
with frequencies of the order of~$\left(m_\text{\tiny env}/d_\text{\tiny Roche}^3 \right)^{\frac{1}{4}} \approx 10^2 \, {\rm keV} \, \left(\tilde L/10^2\right)^{-\frac{3}{4}} \left(m/10^2 M_\odot\right)^{-\frac{1}{2}}$ for the values considered above. 
For comparison, the observed cosmic x-ray and gamma-ray backgrounds range from $10^{-10}\,{\rm W\,m^{-2}\,sr^{-1}}$ at $\sim 10\,{\rm keV}$ to $10^{-13}\,{\rm W\,m^{-2}\,sr^{-1}}$ at $\sim 10\,{\rm GeV}$~\cite{Hill:2018trh}, suggesting that the tidal disruption of these systems could yield photon fluxes within the detectable range for an efficiency factor larger than $10^{-4}$.
This suggests that PBH interpretations of events like GW231123 may be accompanied by an electromagnetic counterpart depending on the properties of the surrounding accretion disks.

\vspace{0.1cm}
\noindent{{\bf{\em Conclusions.}}}
In this work, we have shown that the parameters and merger rate of GW231123, as observed by the LVK Collaboration, are compatible with the PBH hypothesis in the most vanilla scenario, provided that baryonic accretion is sufficiently efficient. In particular, if PBHs experience mildly super-Eddington accretion rates during their late cosmic history, as discussed in Refs.~\cite{Ricotti:2007jk, DeLuca:2020qqa}, their spins are expected to grow according to the standard geodesic prescription for thin-disk accretion, potentially reaching large values. This makes the unique combination of high mass, high spin, and inferred merger rate of GW231123 particularly well-suited to a PBH interpretation, as
they naturally arise within the standard PBH accretion framework~\cite{Byrnes:2025tji}. Our results are robust to variations of the PBH mass function within a plausible range: changes in the central mass and width can accommodate GW231123 while quantitatively affecting the predicted merger rate and the constraints on their abundance. We note, however, that a dedicated assessment of feedback effects is needed~\cite{DeLuca:2023bcr,Serpico:2024cdz,Jangra:2024sif}, as they could reduce the accretion rate or induce thick-disk formation, potentially modifying the resulting spin and mass growth.

A particularly attractive feature of the PBH scenario is its predictive power: future GW observations will be able to thoroughly test this model by probing the existence of further massive binaries with spins following the predicted mass–spin correlation, as well as high-redshift events. This will allow exploration of the putative PBH binary population and its evolution under the effect of accretion.
Interestingly, the parameter space required to explain the inferred GW231123 rate lies near the edge of the exclusion region from non-GW experiments, suggesting that this interpretation could be further tested with GW-independent observations.
Moreover, it is intriguing that accretion may play a crucial role in the interpretation of this event also in astrophysical settings~\cite{Bartos:2025pkv}.

As we have stressed, we have been agnostic on the formation mechanism of the PBHs in the early Universe. In the case in which PBHs are formed due to the collapse of large inflationary overdensities, constraints from the scalar-induced GWs through the NANOGrav15 data provide stringent bounds on the current  PBH abundance in the mass range of the GW231123 event, which, however are significantly relaxed when making use of peak theory to calculate the abundance \cite{Iovino:2024tyg}. This theoretical uncertainty, unless resolved, will limit our ability to constrain the LVK mass range even with future PTA datasets \cite{Cecchini:2025oks}.

Our work can be extended in several directions. It would be interesting to perform an assessment of the PBH interpretation of GW231123 through a more detailed parameter-space exploration of the PBH model, also testing various mass functions. It is worth considering alternative PBH scenarios, such as initially clustered PBHs, which may predict distinct mass functions and merger rates. Finally, a Bayesian comparison between the PBH scenario and astrophysical models explaining massive and spinning binaries in the mass gap, especially hierarchical mergers in dense environments, would be valuable. We leave these aspects to future work.

\vspace{0.1cm}
\noindent{{\bf{\em Acknowledgments.}}}
We thank A. Iovino and G. Perna for interesting discussions.
V.DL. is supported by funds provided by the Center for Particle Cosmology at the University of Pennsylvania. 
A.R. acknowledges support from the Swiss National Science Foundation (Project No. CRSII5 213497) and from the Boninchi Foundation for the project “PBHs in the Era of GW Astronomy”.

\bibliography{draft}

\begin{thebibliography}{143}
\bibitem [{rat()}]{rate}%
  \BibitemOpen
  \href@noop {} {\ }\bibinfo {note} {\hspace{-0.1cm}While the merger rate is
  matched at the observed redshift of GW231123, i.e. $z\sim 0.4$, the bounds on
  the PBH abundance constrain the PBH population at different cosmological
  epochs. To allow for a consistent comparison, we choose to show population
  properties at $z=0$ in Fig.~\ref{fig:main}. Accretion becomes inefficient
  below $z_\text{\tiny cut-off}$ and thus the PBH population does not evolve
  afterward.}\BibitemShut {Stop}%
\bibitem [{dis()}]{disk}%
  \BibitemOpen
  \href@noop {} {\ }\bibinfo {note} {\hspace{-0.1cm}Notice that the condition
  of disk formation relies on the PBH velocity dispersion, and thus varies with
  redshift. Therefore, taking the conservative version of the bound assuming
  spherical accretion is compatible with spin-up at lower
  redshifts.}\BibitemShut {Stop}%
\end{thebibliography}

\end{document}